\documentclass[useAMS,usenatbib]{mn2e}
\usepackage{epsfig,float,natbib,amsmath,longtable,supertabular}

\title[Monthly Notices: \LaTeXe\ guide for authors]
  {Monthly Notices of the Royal Astronomical
  Society: \\ \LaTeXe\ style guide for authors}

  \title[Activity Analysis of Stars in the Southern Hemisphere]
  {An Activity Catalogue of Southern Stars}
\author[J. S. Jenkins et al.]
  {J.S.~Jenkins$^{1}$, H.R.A.~Jones$^1$, C.G.~Tinney$^2$, R.P.~Butler$^3$, \newauthor
 C.~McCarthy$^{4}$, G.W.~Marcy$^{4,5}$, D.J.~Pinfield$^1$, B.D.~Carter$^6$, A.J.~Penny$^{7,8,9}$ \\
  $^1$Centre for Astrophysics Research, University of Hertfordshire, College Lane, Hatfield, Hertfordshire AL10 9AB, email: jsj@star.herts.ac.uk \\
  $^2$Anglo-Australian Observatory, PO Box 296, Epping, NSW 1710, Australia \\
  $^3$Carnegie Institute of Washington, Department of Terrestrial Magnetism, 5241 Broad Branch Road NW, Washington, DC 20015-1305 \\
  $^4$Department of Physics and Astronomy, San Fransisco State University, San Fransisco, CA 94132 \\
  $^5$Department of Astronomy, University of California, Berkeley, CA 94720 \\
  $^6$Faculty of Sciences, University of Southern Queensland, Toowoomba, 4350, Australia \\
  $^7$Rutherford Appleton Laboratory, Chilton, Didcot, Oxon, OX11 0QX, UK \\
  $^8$SETI Institute, 515 North Whisman Road, Mountian View, CA 94043 \\
  $^9$Harvard-Smithsonian Center for Astrophysics, 60 Garden St, MA 02138,USA \\
}

\date{Draft: 05/05}

\pagerange{\pageref{firstpage}--\pageref{lastpage}} \pubyear{2004}

\def\LaTeX{L\kern-.36em\raise.3ex\hbox{a}\kern-.15em
    T\kern-.1667em\lower.7ex\hbox{E}\kern-.125emX}

\begin{document}

\label{firstpage}

\maketitle

\begin{abstract}

We have acquired high-resolution echelle spectra of 225 F6-M5 type stars in the southern hemisphere.  
The stars are targets or candidates to be targets 
for the Anglo-Australian Planet Search.  Ca$\small{\rmn{\textsc{II}}}$~HK line cores were used to derive activity indices for all of these objects.  The indices were 
converted to the Mt.~Wilson system of measurements and log\emph{R}$'_{\rmn{HK}}$ values determined.  A number of these stars had no previously derived activity indices.  
In addition we have also included the stars from \citet{tinney02} using our Mt.~Wilson calibration.  The radial-velocity instability 
(also known as \emph{jitter}) level was determined for all 21 planet-host stars in our dataset.  We find the jitter to be at a level considerably below the radial-velocity 
signatures in all but one of these systems.  19 stars from our sample were 
found to be active (log\emph{R}$'_{\rmn{HK}}$~$>$~-4.5) and thus have high levels of jitter.  Radial-velocity analysis for planetary companions to 
these stars should precede with caution.

\end{abstract}

\section{Introduction}
The Anglo-Australian Planet Search (AAPS) has been monitoring the radial-velocities of over 200 main sequence 
stars since January 1998 (e.g., \citealt{tinney01}).  Radial-velocity searches are the cornerstones of current planet detections, 
giving rise to the bulk of the 171 extrasolar planets (\citealt{butler06}).  
However, dynamical activity in the stellar chromosphere (e.g., \citealt{queloz}; 
\citealt{henry02}; \citealt{paulson04a}) can mimic the radial-velocity signature of a planet.  Indeed, it can 
be the primary source of uncertainty in attributing radial-velocity periodicities to a planetary companion, making the measurement 
of this activity in each planet search target star critical.

The level of activity associated with a target is most commonly determined by measuring the strength of the 
Calcium H and K lines in stellar spectra.  The Mount Wilson HK Project (\citealt{duncan}) has been utilising this technique since 
the mid-1960's for over 100 stars; currently the project monitors over 400 dwarfs and giants.  They have defined a 
log\emph{R}$'_{\rmn{HK}}$ index, which various studies have shown to be a useful indicator of the level
of radial-velocity instability (jitter) in F-K type dwarfs (e.g., \citealt{saar}; \citealt{santos02}; \citealt{wright05}).  
The Vienna-KPNO CaII H\&K Survey (\citealt{strassmeier}) has similar aims. Both these studies target northern hemisphere stars.
There are no such studies conducted in the southern hemisphere that provide the continuous monitoring of activity that these projects 
in the north provide.  Until recently (2006) there was only one large-scale activity analysis of solar-type stars in the southern hemisphere 
and this was provided by \citet{henry} at the Cerro Tololo Inter-American 
Observatory (CTIO).  \citet{gray06} have recently published 
the results of a spectroscopic survey conducted on stars earlier than M0 out to 40pc at the CTIO and the Stewart Observatory.  These 
projects do not provide the long-term monitoring aspect of the Mount Wilson project, yet they provide us with a useful testbed to
the conclusions drawn in the north, whilst also providing an independent sample for statistical analysis.   

When the AAPS list was drawn up, the vast 
majority of targeted stars were not on the activity list compiled by \citet{henry}.  Henry et al. acknowledge that their primary sample of 650 stars south of 
-25$^{\rmn{o}}$ only comprise around 50\% of the total number of solar-type stars in this region of the southern sky down to V$\sim$9.  Hence, to measure the chromospheric 
Ca$\small{\rmn{\textsc{II}}}$~HK core emission for all AAPS targets not in the Henry et al. catalogue we have taken spectra of 225 stars from the AAPS, a number of which 
had no previously derived activity indices.  Also the majority of these objects are already being monitored by the AAPS for the presence of planetary companions, however 
some are being scrutinised as potential target candidates for an expansion of the target list.

\section{Observations and Reduction}
Six nights of observations were taken over a period of four years and are listed in Tables~\ref{tab:calibrators} and \ref{tab:activity}. The observations were 
made using the University College London 
Echelle Spectrograph mounted on the 3.9-m Anglo-Australian Telescope (AAT), and followed the observing procedures described by \citet{tinney02}.  
We used the EEV2 2048 x 4096 13.5-$\micron$ pixel detector, with a 
Ca$\rmn{\small{\textsc{II}}}$ HK quantum efficiency (QE) of $\sim$65\%, for observations made on the nights 2001 August 04, 2002 July 20, 2003 July 21 and 2005 June 16.  
The CCD was spatially 
binned by two to give effective slit lengths of 11, 14, 23.5 and 17.5 pixels.  The dispersion at the Ca$\rmn{\textsc{II}}$~HK lines is 0.02\AA\ pixel$^{-1}$, 
giving resolutions of 3.5, 4, 6 and 4 pixels or 0.07, 0.08, 0.12 and 0.08\AA.  The remaining nights of observations were conducted on 2004 August 23 and 24 
using the MIT and Lincoln Labs (MITTL3) 2048 x 4096 
chip.  The chip has relatively low QE ($\sim$18\%) around the HK line region, however as these are bright sources this presented no problems.  The CCD setup was similar 
to the EEV2 chip giving effective slit lengths of 23.5 pixels.  The 
dispersion was again similar to that of the EEV2 runs (0.02\AA\ pixel$^{-1}$) giving a resolution of 4.5 pixels or 0.09\AA.

\begin{figure}
\vspace{4.5cm}
\includegraphics{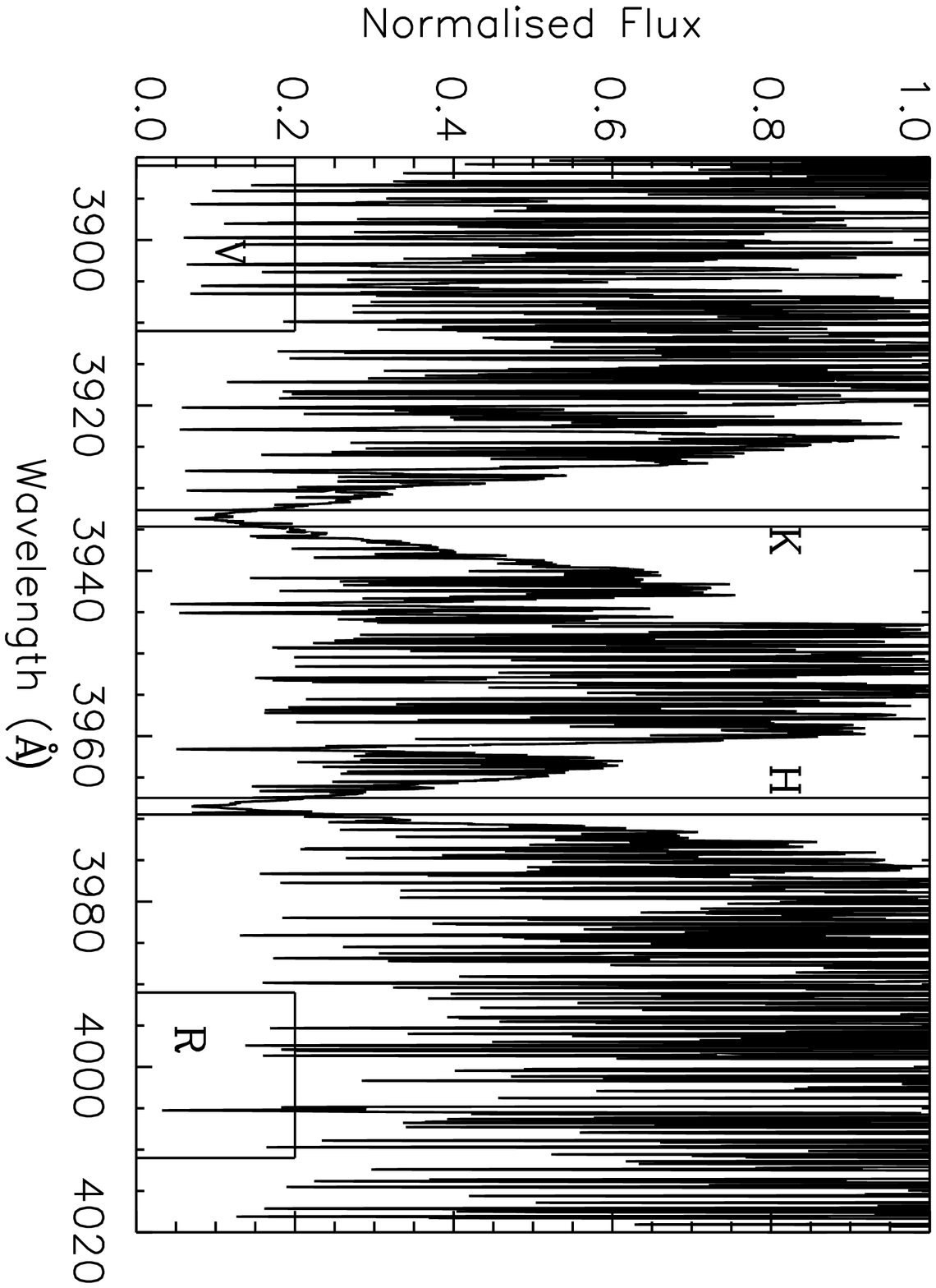}
\vspace{5cm}
\includegraphics{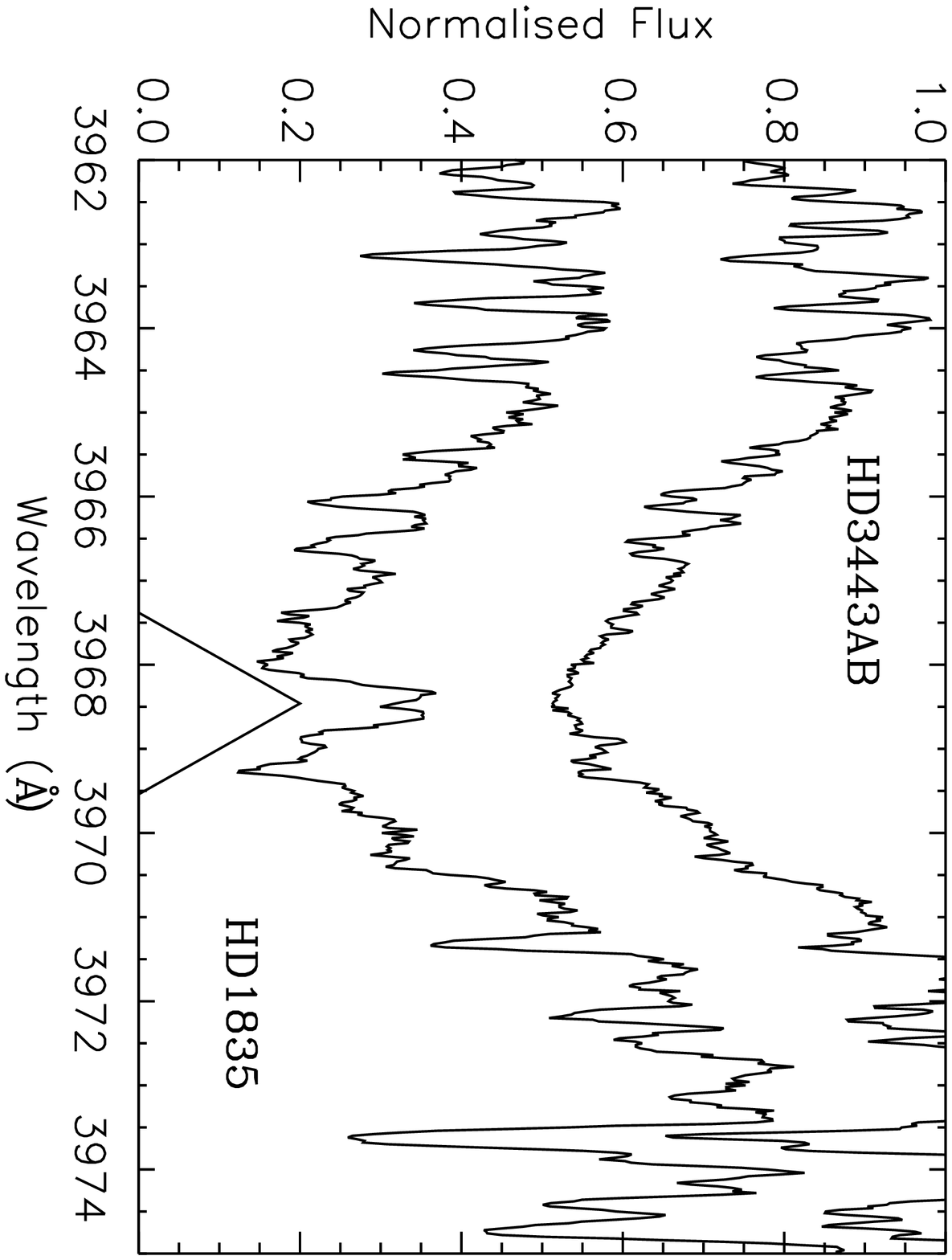}
\vspace{0.8cm}
\caption{The Calcium H and K emission features, obtained at the AAT, for the star HD190248 (top panel).  The R, V, H and K bandpasses are indicated 
for reference.  The narrow features, other than H and K, are due to weak absorption lines not noise.  The bottom panel shows 
the triangular bandpass used to integrate the 
flux for the Ca$\rmn{\textsc{II}}$~H feature.  The calibration stars shown are HD3443AB (inactive: log\emph{R}$'_{\rmn{HK}}$=-4.89) and HD1835 
(active: log\emph{R}$'_{\rmn{HK}}$=-4.45).  Both have been offset from zero for clarity.}
\label{spectra}
\end{figure}

225 F6-M5 dwarfs and sub-giants were observed in seeing ranging from 0.7-1.4 arcseconds through slits of sizes 1.0~x~3.5, 1.0~x~4.0 and 1.0~x~6.0 
arcseconds.  We have also included the observations in \citet{tinney02} to complete this catalogue.  Exposure times ranged from 30 
seconds to 900 seconds for the faintest objects. This provided signal-to-noise ratios per 0.02\AA\ wavelength pixel of between 10 and 80.  The target stars were 
taken from the AAPS target list and were supplemented by calibration stars taken from the Mount Wilson HK Project (\citealt{duncan}).  These are used 
to calibrate onto the Mount Wilson system of measurements.  All 
calibration stars are listed in Table 1, along with their visual magnitudes, colours, spectral types and derived activity indices.

The reduction was accomplished using standard \sc echomop \rm protocols (\citealt{mills96}).  Prior to the \sc echomop \rm 
procedure the files were prepared by performing a bias or overscan subtraction and rotating them to \sc echomop's \rm desired alignment.  
Each of the five orders (142~-~146) were then traced, clipped, flat fielded and the scattered light was removed.  No sky subtraction 
was needed as the sky brightness around the CaHK lines is negligible compared to the brightness of our sources.  Wavelength calibration was 
performed using ThAr arc spectra acquired for this purpose.  
The data acquired using the MITTL3 chip were then flux calibrated, as were the data from \citet{tinney02}, in order to correct the blaze and inter-order sensitivity.  
This was accomplished using a measurement of the standard $\mu$~Col (\citealt{turnshek}).  
None of the other three nights were flux calibrated.  All spectra were then scrunched to a linear scale with bin widths of 0.02\AA.  All stars were then 
cross-correlated with an observation of HD216435 and the barycentric velocity for this star was applied, leaving a series of zero velocity spectra.  \citet{jones02} have 
shown that HD216435 has a radial-velocity variation of 20ms$^{-1}$ over a period of 1326~days (3.7~years).  This is negligible when compared with the 
barycentric correction.  The spectra were then normalised to the continuum region between 3991 
and 4011\AA, which represents the outer bandpass region (R) we used to acquire our desired activity index (\emph{S$_{\rmn{AAT}}$}).

To determine robust CaHK activity indices we employed four bandpasses centred around the calcium HK line cores.  The V and R 
bandpasses both have square profiles with widths of 20\AA\ and are centred on the continuum at 3901\AA\ and 4001\AA\ respectively.  The 
K and H bandpasses have triangular profiles with FWHM's of 1.09\AA\ and are centred on the cores themselves, at 3933.667\AA\ 
and 3968.470\AA\ respectively.  Fig.~\ref{spectra} (upper) shows the final spectrum after the reduction procedure for the star HD190248.  
All four passbands are highlighted for 
reference.  The H and K line cores are clearly evident.  This star has a spectral type of G5IV-V.  We find this star to be chromospherically quiet, 
with a final log\emph{R}$'_{\rmn{HK}}$ of -5.03.  The lower plot in Fig.~\ref{spectra} highlights the triangular bandpasses used to integrate the flux 
of the H line core.  Two calibration stars are shown for reference, those are HD3443AB (top) and HD1835 (bottom), and both have been offset from zero for clarity.  
The stars are both main sequence stars and have spectral types of K1V and G3V respectively.  The more active of the two (HD1835) exhibits some marked emission 
at the central core, whereas the inactive star (HD3443AB) has a deep central minima.  It is these characteristics that allow accurate activity indices to be generated.

\section{ANALYSIS}

\begin{figure}
\vspace{4.5cm}
\hspace{-4.0cm}
\includegraphics{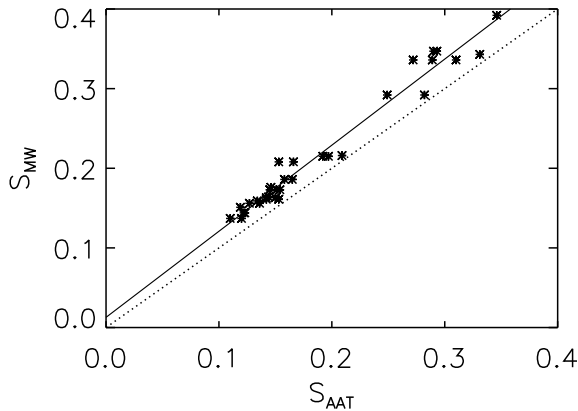}
\vspace{0.6cm}
\caption{Calibration of the \emph{S}$_{\rmn{AAT}}$ index onto the \emph{S}$_{\rmn{MW}}$ system of measurements.  The solid line represents a linear least-squares fit 
described by \emph{S}$_{\rmn{MW}}$=(1.081\emph{S}$_{\rmn{AAT}}$)+0.013, with a total RMS scatter of 0.014.  The dotted line shows a 1:1 relationship.}
\label{s_cal}
\end{figure}

\subsection{\emph{S} Indices}

Our spectra have been used to measure Ca$\rmn{II}$ HK emission (\emph{S$_{\rmn{MW}}$}) indices (\citealt{duncan}). This index is a measure of the ratio between the integrated 
flux in two triangular bandpasses
(with FWHM's of 1.09\AA)  centred on the Ca$\rmn{II}$ H (3968.470 \AA) and K (3933.664 \AA) lines, against the integrated flux in two square 
20\AA\ continuum bandpasses at either side of the HK features, centred at 3901\AA\ (V) and 4001\AA\ (R). The Mt. Wilson (MT) Project 
determine this index individually each night using a specialised multi-channel photometer. \emph{S}$_{MW}$ is defined as:

\begin{equation}
\label{eq:smw}
\emph{S}_{\rmn{MW}} = \alpha\frac{N_{\rmn{H}} + N_{\rmn{K}}}{N_{\rmn{R}} + N_{\rmn{V}}}
\end{equation}

where \emph{N$_{i}$} is the number of counts in each bandpass (where \emph{i}=H,~K,~V~and~R) and $\alpha$ is a constant that is determined 
each night by the observation of standards.

Using the same methodology used to determine Eq.~\ref{eq:smw}, \citet{tinney02} have shown that the emission index for stars observed at the AAT 
(\emph{S}$_{\rmn{AAT}}$) is given by:

\begin{equation}
\label{eq:saat}
S_{\rmn{AAT}} = \frac{N_{\rmn{H,1}} + N_{\rmn{H,2}} + N_{\rmn{K,1}} + N_{\rmn{K,2}}}{2(N_{\rmn{R}} + N_{\rmn{V}})}
\end{equation}

Here the \emph{N$_{\rmn{HK}}$} terms have two components.  This is due to positioning of the echellogram, which causes
both the H and K features to appear in two adjacent orders. 

We used Eq.~\ref{eq:saat} to combine the integrated flux values from all bandpasses, giving us a set of \emph{S}$_{\rmn{AAT}}$ values we could use 
to calibrate onto the MW system of measurements.  We have employed slightly different instrumental setups in our study and have observed our targets across a range of S/N 
ratios therefore we have included individual measurements for multiple objects in our calibrations.  We have also chosen the most stable calibrators possible from 
\citet{duncan}, allowing us a 
better estimation of the variability error in our final activity index.  Columns 3 and 4 of \citet{duncan} give the minimum and maximum \emph{S} indices obtained at 
Mt.~Wilson.  Over the course of one observing season the measured \emph{S} index can vary on the level of $\pm$0.05, therefore selecting the most stable calibrators helps 
us reduce this uncertainty.  Fig.~\ref{s_cal} shows the least-squares fit used to calibrate onto \emph{S}$_{\rmn{MW}}$.  The 
calibration we employ (\emph{S}$_{\rmn{AAT}}$) has a slope of 1.081$\pm$0.001 and a zero-point offset of 0.013$\pm$0.007.  The RMS scatter about the fit is 0.014.  
This level of scatter is lower than the scatter found by \citet{tinney02} who also calibrated against stars from the CTIO study.  
\citet{henry} employed a different approach to both the MW Project and our AAT analysis.  They centred 4\AA-wide square bandpasses on the HK line cores, 
compared to the 1.09\AA\ triangular bandpasses employed at MW.  When the bandpasses here are widened non-linear 
calibrations are needed to convert to the MW system of measurements.  We suspect this leads to increased systematic errors in HK values.  Indeed when we include all objects 
in this study with CTIO indices in our calibration we find a similar scatter to \citet{tinney02} of around 0.02.

Table~\ref{tab:activity} shows all derived \emph{S}$_{\rmn{AAT}}$ activity indices, after calibration onto the MW system of measurements, along with their 
one sigma photon-counting uncertainties.  The table is split by epoch and there are a few objects with more than one measurement.  In the majority of 
cases the photon-counting errors are an order of magnitude below that of the overall scatter from the calibration.  In order to fully quantify all sources of error in 
the reduction procedure, such as scattered light removal, proper blaze removal, precise flux calibration etc, we used the stable star HD10700 ($\tau$~Ceti) as a proxy 
for the reduction errors.  We have acquired three separate measurements of $\tau$~Ceti spanning a period of three years and due to the extremely small variation of $\sim$1\% 
(\citealt{baliunas}) this represents a good indicator of the random errors in the reduction procedure.  The standard deviation of the three measurements here is 3\%, which 
is slightly lower than that of the Keck and Lick errors from \citet{wright04}.  However, we only have three separate measurements, which likely gives rise to a slightly 
lower estimation of the random errors.  This 3\% error should also be taken into account when quantifying the significance of any of the activity indices in 
Table~\ref{tab:activity}.

\subsection{log\emph{R$'_{\rmn{HK}}$} Values}

\begin{figure}
\vspace{4.5cm}
\hspace{-4.0cm}
\includegraphics{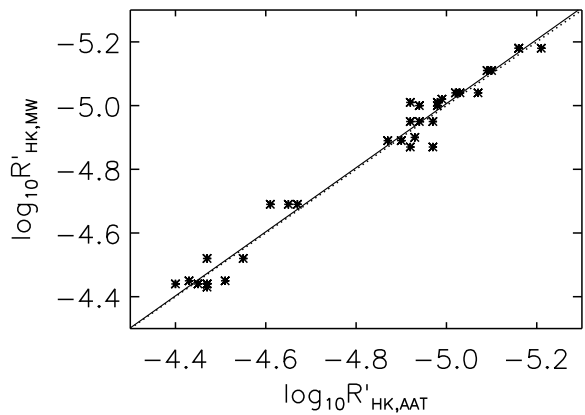}
\vspace{1.4cm}
\caption{Derived log\emph{R}$'_{\rmn{HK}}$ chromospheric activities for the calibration stars on the AAPS target
list.  The solid line is a least-squares fit with a slope of 1.006 and a zero-point offset of 0.023.  This result highlights the close relationship between 
the AAT and MW measurements.  The dotted line is a direct 1:1 relationship.}
\label{rhk}
\end{figure}

The \emph{S} index provides an estimate of both the photospheric and chromospheric flux, and hence a combined Ca$\rmn{\textsc{II}}$ emission feature.  In order to 
obtain actual activity values we have to concentrate on chromospheric flux, meaning photospheric effects have to be removed.  This is accomplished by 
normalising the chromospheric emission to the bolometric luminosity of the star, and is most commonly designated by the log\emph{R}$'_{\rmn{HK}}$ parameter.  
We have followed \citet{noyes} to convert all main sequence star values from \emph{S$_{\rmn{AAT}}$} to
log\emph{R}$'_{\rmn{HK,AAT}}$.  We extrapolated the Noyes et al. conversion to include later spectral-types due to the lack of information on these stars.  The Noyes 
et al. calibration is intended for stars in the colour range of B-V$\sim$0.44-0.90, which means the final log\emph{R$'_{\rmn{HK}}$} measurements of stars 
redder than 0.90 have a larger uncertainty than those in the calibration range. 
The derived values for the calibration stars are given in Table~\ref{tab:calibrators}, along with the Duncan et al. 
log\emph{R}$'_{\rmn{HK,MW}}$ values.  The values for all other objects are listed in Table~\ref{tab:activity}.  The Hipparcos 
B-V and visual magnitudes are shown in both tables for reference, and to highlight the photometric contribution to the \emph{S}$_{\rmn{AAT}}$ values.

In Fig.~\ref{rhk} we have plotted our log\emph{R}$'_{\rmn{HK,AAT}}$ values against the published values from Duncan et al.  It can be seen from the gradient of the fit 
that there is good agreement with the MW data points.  The gradient of the linear trend is 1.006$\pm$0.001 with an offset of 
0.023$\pm$0.137.  The RMS of the fit is 0.041, highlighting just how tight our values are to the published Mt.~Wilson data.  By using all the measurements in the calibration, 
and not just the means, we allow the error to include any variability and systematic effects.  The dotted line 
in the figure shows a direct 1:1 relationship and it can be seen how close the log\emph{R}$'_{\rmn{HK,MW}}$ and log\emph{R}$'_{\rmn{HK,AAT}}$ are.  We therefore employ 
no further calibration.  Due to the inherently variable nature of stellar activity and errors in the reduction procedure, we believe the deduced RMS is a better measure 
of the overall error budget in each individual measurement.

\section{Discussion}

Table~\ref{tab:activity} includes new and updated log\emph{R}$'_{\rmn{HK}}$ activities for 21 planetary systems from the AAPS.  It must be noted that many more stars on 
this list probably have planetary systems that are still awaiting discovery or are below the detectability threshold of current radial-velocity surveys.    We also include 
the observations from \citet{tinney02} using our new MW activity calibration.  Most of these planetary hosts had 
previously derived log\emph{R}$'_{\rmn{HK}}$ values (e.g., \citealt{henry}; \citealt{wright04}; \citealt{saffe05}).  \citet{gray06} have recently completed the analysis of 
stars in the southern hemisphere for the NStars Project and the values in here are mostly in agreement with this work.  However, the agreement with this work is not as 
tight as with \citet{wright04}, most likely due to the different setups employed in this study as the Gray et al. approach used 4\AA\ bandpasses, as opposed to the 1.09\AA\ 
used in this work and the MW project.  Indeed when we increase the widths of our bandpasses here, non-linear calibrations are needed onto the MW system.  

Knowledge of the level of activity induced jitter in any radial-velocity 
measurement is essential to quantify the errors on a detected planetary fit.  Jitter can vary on a timescale of days due to a 
number of different factors such as magnetic flux tube evolution, sub-photospheric convection, stellar oscillations and surface rotation of spots (\citealt{marcy05a}) 
and can thus have a significant bearing on radial-velocity measurements.  Therefore determining the jitter level of independent measurements allows one to generate a more 
precise velocity point and hence a more precise overall Keplerian fit.

The 21 planetary systems in this work have log\emph{R}$'_{\rmn{HK}}$ values ranging from -5.35 (HD27442) to -4.43 (HD22049).  HD22049 (aka. Eps~Eri) already had measurements 
taken at MW and they agree with our derived mean value of -4.43, confirming the high activity of the host star.  \citet{saar98}, \citet{santos00} and \citet{wright05}
have carried out studies probing the relationship between stellar activity and radial velocity jitter ($\sigma_{rv}'$).  Wright provides an empirical estimate 
of jitter, culminating in a jitter metric ($\sigma'_{\rmn{rv}}$), which is a function of stellar evolution (for a full detailed description of the analysis, see Eqs.~1~-~7 
Wright 2005).  Applying this methodology, and using Hipparcos B-V, M$_{\rmn{V}}$ and a T$_{\rmn{EFF}}$ from \citet{valenti05}, we find the jitter level for the most active 
planet-host star in our catalogue (HD22049) 
to be $\sim$5.7ms$^{-1}$ with 20$^{\rmn{th}}$ and 80$^{\rmn{th}}$ percentiles of 4.6 and 9.5~ms$^{-1}$ respectively (see Wright appendix for an explanation of 
percentiles).  This low level of jitter for such an active star arises because the star is spectral type K.  Jitter falls off at later spectral types 
allowing lower radial-velocity trends to be found around active K stars than F stars.  The radial-velocity amplitude (K) for HD22049 is 19$\pm$1.7ms$^{-1}$ 
\citep{hatzes00}.  This level of velocity is significantly above our derived jitter.  We therefore conclude that activity is not the 
source of the radial-velocity signature for HD22049.

Two other planet-host stars are shown to be quite active, both have log\emph{R}$'_{\rmn{HK,AAT}}$ values larger than Eps~Eri.  The first of these is HD13445 and it has a 
log\emph{R}$'_{\rmn{HK,AAT}}$ of -4.64.  As this star is a K-dwarf the level of jitter it exhibits is not extremely high but it should be in the range 6-13ms$^{-1}$.  The 
planetary fit to this star has an amplitude of 380$\pm$1ms$^{-1}$ with a period of 15.78 days and from this a minimum mass of $\sim$4M$_{\rmn{J}}$ was derived for the planet 
(\citealt{queloz00}).  It is clear the jitter for this star is significantly 
lower than the observed radial-velocity measurements.  The second of the two stars is HD17051 ($\iota$~Hor) and it has a log\emph{R}$'_{\rmn{HK,AAT}}$ value of -4.59.  
\citet{kurster00} announced the detection of a planet with a minimum mass of 2.26M$_{\rmn{J}}$ and a period of 320.1$\pm$2.1~days.  The amplitude of the planetary fit 
is 67ms$^{-1}$ with an RMS scatter of 27ms$^{-1}$.  The internal errors were estimated to be 17ms$^{-1}$ and they speculated the difference was activity induced, with 
a jitter of 20ms$^{-1}$ making up this difference.  \citet{butler01} obtained an RMS of 10.4ms$^{-1}$ for this star using measurements made at the AAT and \citet{saar98} 
find a jitter of 10ms$^{-1}$, consistent with the AAT scatter.  A further 10 measurements have been taken by Bulter et al. since 2001 over a period of 4~years and the 
current best-fit single-planet Keplerian has an RMS of 20ms$^{-1}$, consistent with the results from \citet{kurster00}.  We find a jitter range of 9-19ms$^{-1}$, which 
agrees well with observations and the higher end of this range can go a long way toward explaining the high level of scatter observed by \citet{kurster00} and Butler et al. 
(private communication).

17 out of the other 18 remaining planet-hosts objects are relatively inactive.  These are HD142, HD2039, HD20782, HD23079, HD27442, HD30177, HD70642, 
HD73526, HD76700, HD102117, HD134987, HD142415, HD154857, HD169830, HD179949, HD216435 and HD216437.  We find that the derived jitter value can not explain the observed 
stellar radial-velocity signal confirming the planetary hypothesis.  However, the remaining star (HD10647) is spectral type F and we find the radial-velocity 
jitter is likely to be similar to the reported planetary signal.  \citet{mayor03} announced a best-fit Keplerian to the data with a K of 18$\pm$1ms$^{-1}$, 
giving rise to a planet with 
a period of 1040~days (2.85~yrs) and a Msin\emph{i} of $\sim$0.91~M$_{\rmn{J}}$.  We find the log\emph{R}$'_{\rmn{HK,AAT}}$ index to be -4.70.  This relates to a jitter 
of $\sim$11-23ms$^{-1}$ at the 1-sigma level.  This level of jitter is similar to the 
derived planetary signature.  The star also exhibits a strong IR excess (\citealt{decin}) which may be indicative of the presence of a disk.  This also agrees 
well with the active nature of the star as this indicates the star is young.  Mayor acknowledged the jitter phenomenon and by the use of a bisector analysis, 
they found no evidence for any periodic line profile variability.  \citet{jones04} find only weak evidence for a planetary companion in the AAPS dataset for this object.  
However, four additional measurements have been taken at the AAT for this star over a period of 2~years.  The current best-fit Keplerian has a period of 2.73~years with an 
amplitude of 17.9ms$^{-1}$.  The RMS to the fit is 8.9ms$^{-1}$.  This signature is in agreement with the signal announced by \citet{mayor03}, however with a false alarm 
probability of 0.25 more data points will be needed to obtain robust planetary parameters.

\subsection{log\emph{R}$'_{\rmn{HK}}$ Distribution:-}
\begin{figure}
\vspace{4.5cm}
\hspace{-4.0cm}
\includegraphics{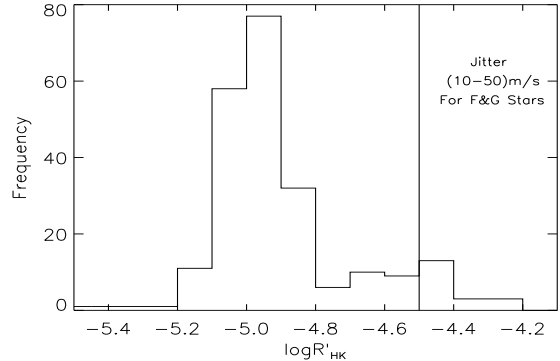}
\vspace{0.5cm}
\caption{A histogram of the log\emph{R}$'_{\rmn{HK}}$ activity index distribution for the stars on the AAPS target list.  It is clear 
that the majority of the stars observed by the AAPS are inactive, with the peak residing at values below -4.90.  The solid vertical line bounds 
the region inhabited by very active stars.  The lower limit radial-velocity jitter range is shown for F and G stars in this region.}
\label{rhk_dis}
\end{figure}

Fig.~\ref{rhk_dis} shows the distribution of stellar activity for 225 stars on the AAPS target list.  The selection criteria for the list 
itself targets more evolved stars, and as such the inactive peak was expected.  We see the indication of a possible binomial distribution in activity with peaks 
centred at -4.95 and -4.45.  This result has been found before in both the northern and southern hemispheres (e.g., \citealt{vaughan}; \citealt{henry}).  

There are a  number of highly active (log\emph{R}$'_{\rmn{HK}}~>$~-4.5) 
stars on this list.  For the assessment of highly active stars we use the \citet{santos00} jitter relation (see  Eqs.~2~-~4 from \citealt{santos00}), as the Wright 
model is out of range for some of our active stars.  The vertical solid line in 
Fig.~\ref{rhk_dis} represents the lower boundary of these active stars.  There are 19 stars 
in this region.  Since the \emph{R}$'_{\rmn{HK}}$-$\sigma'$(Vr) relation decreases with 
spectral type, the jitter values of half of these are closer to the lower end of this range, as 9 of these stars are K-dwarfs.  However, 
all are expected to have jitter values in excess of $\sim$7ms$^{-1}$, therefore the low-amplitude radial-velocity signatures of companions must be 
treated with care.  Also two of the active set are M-dwarfs, and with B-V colours $>$1.5 they are outside the model limits for jitter determination.  \citet{wright05} 
shows that M-type stars statistically exhibit higher jitter levels than F,G,K-type stars.  Therefore, we expect these two active M stars to exhibit high levels of 
jitter.

Analysis of the remaining F and G-type stars, HD1835, HD11131, HD13246, HD19632, HD88201, HD90712, HD129060 and HD152391 show these are 
likely to have jitter values around $\sim$10-30ms$^{-1}$.  HD1835 and HD11131 were used as 
calibrators, and both the MW and CTIO studies have found these to be highly active. 
This level of jitter has already proven sufficient to mimic a 
planetary companion (e.g., \citealt{queloz}; \citealt{henry02}).  It must be noted that the log\emph{R}$'_{\rmn{HK}}~>$~-4.5 is a somewhat arbitrary 
cut-off.  An F-type star, such as HD16673, with a log\emph{R}$'_{\rmn{HK}}$ value of -4.64, still provides a jitter range of around: 
$\sigma\sim$11-24ms $^{-1}$.  Therefore, it is necessary to take great 
care when dealing with radial-velocity amplitudes for all F-type stars $>$-4.70.  Indeed 
the two active F-type stars (HD13246 and HD129060) exhibit extremely high levels of jitter.  Having log\emph{R}$'_{\rmn{HK}}$ values of -4.38 and 
-4.41 respectively, they exhibit jitter of around $\sim$17-40ms$^{-1}$.  This level of jitter puts serious limitations on the potential for planet discoveries 
around these stars.  The AAPS baseline is 8~years and even giant planets with orbital periods longer than this do not generate large radial-velocity amplitudes.  
For example, Jupiter, with a semimajor axis of 5.2AU, induces a radial-velocity motion of $\sim$12ms$^{-1}$ on the Sun over a 12~year period.  The jitter 
level exhibited by both these stars could serve to mask the planetary signature of an orbiting Jupiter even once the baseline is sufficient to detect them.

Due to the variable nature of stellar activity, in particular when dealing with active 
stars, it is important to obtain multi-epoch activity indices.  This will allow planet searches to state categorically that the jitter-level 
is significantly lower than any companion's periodic radial-velocity signature.  It is even more essential in the southern hemisphere, where no such 
program yet exists.  We have obtained more than one measurement for a small number of our stars, but with a maximum of three observations, we are still limited 
by a lack of data.  At MW a number of measurements are obtained of the same object over an
observing season.  The difference in log\emph{R}$'_{\rmn{HK}}$-index over 
one season can change at the level of $\sim\pm$0.2 (see \citealt{duncan} Table 1 column 5).  
This variation was recorded using a single well defined 
setup  (e.g., HD16673: 202 measurements: -4.66~$\le$~log\emph{R}$'_{\rmn{HK}}$~$\le$~-4.82).  Another example would be our own parent star.   
The log\emph{R}$'_{\rmn{HK}}$-index of the Sun varies from its typical value of $\sim$-5.10 to 
-4.96 during the Solar minimum (\citealt{wright04}).  This level of discrepancy highlights the difficulty when attempting to calibrate between systems and when trying to 
fully quantify activity.  This is why gaining multiple measurements using a well defined standard setup is essential.

Time series analysis also allows other avenues of study to be undertaken.  Stellar ages and rotation periods can be empirically estimated 
using log\emph{R}$'_{\rmn{HK}}$ relationships.  For instance the age-chromospheric activity relation of \citet{soderblom}, most recently updated to its current form 
by \citet{baliunas} and taken from \citet{wright05} is given by:

\begin{equation}
\label{eq:age}
log[t(\rmn{yr})] = -0.0522R^3_5+0.4085R^2_5-1.334R_5+10.725
\end{equation}

here \emph{R}$_5$=10$^5$\emph{R}$'_{\rmn{HK}}$ and \emph{R}$'_{\rmn{HK}}$ represents the activity level for a solar-type star averaged over many stellar 
activity cycles.  The more observations one obtains, the closer one gets to the actual age derived from this relationship.  This is because taking multiple 
measurements allows a better determination of the mean activity index for main sequence stars.  Using the Sun's variability range from above ($\sim$-5.10 
to -4.96) causes measurements of the solar age to differ by several Gyrs (e.g. 7.92~-~4.93~Gyrs).  
Variability knowledge is essential to obtain reliable age and rotation estimates of stars.

\begin{table*}
\caption{\emph{S} and log\emph{R$'_{\rmn{HK}}$} activity values for all stars used to calibrate onto the Mt.~Wilson system of measurements.  \underline{Column Headings:-} 
HD Number:- Henry Draper catalogue identifiers.  Johnson B-V, V and Spec Type all taken from Hipparcos.  
\emph{S}$_{\rmn{AAT}}$ \emph{after} calibration onto the Mt.~Wilson 
system.   \emph{S}$_{\rmn{MW}}$ is the Mt.~Wilson index used to perform this calibration, taken from column 5 of Table 1 and 
columns 4-7 of Table 3 from \citet{duncan}.  log\emph{R}$'_{\rmn{HK,MW}}$ are derived from \citep{duncan}.  Note: the photon-counting errors are indicative only, a random 
error of $\ga$~3\% must be taken into account in all instances.  This error was estimated from measurements of $\tau$~Ceti.}. \label{tab:calibrators} 
\
\begin{tabular}{cccccccc}
\hline 
\multicolumn{1}{c}{HD} & \multicolumn{1}{c}{B-V} & \multicolumn{1}{c}{V} & \multicolumn{1}{c}{Spec Type} & \multicolumn{1}{c}{\emph{S}$_{\rmn{AAT}}$}
& \multicolumn{1}{c}{\emph{S}$_{\rmn{MW}}$} & \multicolumn{1}{c}{log\emph{R}$'_{\rmn{HK,AAT}}$} & \multicolumn{1}{c}{log\emph{R}$'_{\rmn{HK,MW}}$} \\ \hline

                \underline{Calibrators} &        &       &       &    &    &   &                        \\
HD1835 &  0.659 &   6.39 &        G3V &  0.330$\pm$0.002, 0.326$\pm$0.002, 0.371$\pm$0.002  &  0.347 &  -4.47, -4.47, -4.40 &  -4.44 \\
HD3443AB &  0.715 &   5.57 &        K1V &  0.184$\pm$0.001, 0.191$\pm$0.001 &  0.186 &  -4.87, -4.90 &  -4.89 \\
HD3795 &  0.718 &   6.14 &     G3/G5V &  0.150$\pm$0.001, 0.157$\pm$0.001, 0.160$\pm$0.001 &  0.156 &  -5.02, -5.03, -5.07  &  -5.04 \\
HD9562 &  0.639 &   5.75 &       G2IV &  0.145$\pm$0.001, 0.146$\pm$0.001 &  0.144 &  -5.09, -5.10 &  -5.11 \\
HD10700 &  0.727 &   3.49 &        G8V &  0.169$\pm$0.001, 0.176$\pm$0.001, 0.179$\pm$0.001  &  0.173 &  -4.92, -4.94, -4.97 &  -4.95 \\
HD11131 &  0.654 &   6.72 &         G0 &  0.307$\pm$0.002, 0.325$\pm$0.002, 0.348$\pm$0.002 &  0.336 &  -4.43, -4.47, -4.51 &  -4.45 \\
HD16673 &  0.524 &   5.79 &        F6V &  0.221$\pm$0.001, 0.225$\pm$0.001, 0.239$\pm$0.002 &  0.215 &  -4.61, -4.65, -4.67  &  -4.69 \\
HD23249 &  0.915 &   3.52 &       K0IV &  0.132$\pm$0.001, 0.142$\pm$0.001, 0.142$\pm$0.001 &  0.137 &  -5.16, -5.16, -5.21 &  -5.18 \\
HD26965 &  0.820 &   4.43 &        K1V &  0.178$\pm$0.001, 0.192$\pm$0.001 &  0.208 &  -4.92, -4.97  &  -4.87 \\
HD30495 &  0.632 &   5.49 &        G3V &  0.282$\pm$0.002, 0.317$\pm$0.002 &  0.292 &  -4.47, -4.55 &  -4.52 \\
HD81809 &  0.642 &   5.38 &        G2V &  0.170$\pm$0.001 &  0.160 &  -4.93 &  -4.90 \\
HD115617 &  0.709 &   4.74 &        G5V &  0.166$\pm$0.001, 0.178$\pm$0.001 &  0.161 &  -4.92, -4.98 &  -5.01 \\
HD152391 &  0.749 &   6.65 &        G8V &  0.387$\pm$0.003, 0.458$\pm$0.003 &  0.392 &  -4.36, -4.45 &  -4.44 \\
HD158614 &  0.715 &   5.31 &     G8IV-V &  0.167$\pm$0.001, 0.175$\pm$0.001 &  0.163 &  -4.94, -4.98 &  -5.00 \\
HD219834AC &  0.787 &   5.20 &    G6/G8IV &  0.169$\pm$0.001 &  0.164 &  -4.99 &  -5.02 \\

\hline
\end{tabular}
\medskip
\end{table*}

\begin{table*}
\caption{\emph{S$_{\rmn{AAT}}$} and log\emph{R}$'_{\rmn{HK,AAT}}$ for stars on or under consideration for the AAPS.  \underline{Column Headings:-} HD Number:- Henry Draper 
catalogue identifiers (numbers with associated asterisks (*) highlight planet bearing stars).  Johnson B-V, V and Spec Type all taken from the Hipparcos catalogue.  \emph{S}$_{\rmn{AAT}}$:- the AAT activity index \emph{after} calibration onto the Mt. Wilson 
system of measurements.  For objects with multiple measurements we have included all the individual indices.  
log\emph{R}$'_{\rmn{HK,AAT}}$:- final activity values generated 
following the \citet{noyes} methodology.  Note: the photon-counting errors are indicative only, a random error of $\ga$~3\% must be taken into account in all instances.  
This error was estimated from measurements of $\tau$~Ceti.}. \label{tab:activity}
\
\begin{tabular}{cccccc}
\hline 
\multicolumn{1}{c}{HD}& \multicolumn{1}{c}{B-V}& \multicolumn{1}{c}{V} & \multicolumn{1}{c}{Spec Type}& \multicolumn{1}{c}{\emph{S}$_{\rmn{AAT}}$} 
& \multicolumn{1}{c}{log\emph{R}$'_{\rmn{HK,AAT}}$} \\ \hline

                \underline{Measurements}                &       &            &                        &      &  \\

\underline{2001 August 04$^{\rmn{th}}$} &        &       &       &    &                       \\
*HD142 &  0.519 &   5.70 &       G1IV &  0.161$\pm$0.001 &  -4.95 \\
*HD2039 &  0.656 &   9.00 &  G2/G3IV/V &  0.180$\pm$0.001 &  -4.89 \\
HD2587 &  0.748 &   8.46 &        G6V &  0.152$\pm$0.001 &  -5.07 \\
HD3823 &  0.564 &   5.89 &        G1V &  0.160$\pm$0.001 &  -4.97 \\
HD6735 &  0.567 &   7.01 &        F8V &  0.173$\pm$0.001 &  -4.89 \\
HD7199 &  0.849 &   8.06 &     K0IV/V &  0.177$\pm$0.001 &  -4.99 \\
HD7570 &  0.571 &   4.97 &        F8V &  0.166$\pm$0.001 &  -4.93 \\
HD9280 &  0.760 &   8.03 &         G5 &  0.148$\pm$0.001 &  -5.09 \\
HD10180 &  0.629 &   7.33 &        G1V &  0.165$\pm$0.001 &  -4.96 \\
HD10647 &  0.551 &   5.52 &        F8V &  0.213$\pm$0.001 &  -4.70 \\
HD11112 &  0.637 &   7.13 &        G4V &  0.158$\pm$0.001 &  -5.00 \\
*HD13445 &  0.812 &   6.12 &        K0V &  0.302$\pm$0.002 &  -4.64 \\
HD16417 &  0.653 &   5.78 &        G1V &  0.148$\pm$0.001 &  -5.08 \\
*HD17051 &  0.561 &   5.40 &       G3IV &  0.249$\pm$0.002 &  -4.59 \\
HD18907 &  0.794 &   5.88 &     G8/K0V &  0.147$\pm$0.001 &  -5.11 \\
HD19632 &  0.678 &   7.29 &     G3/G5V &  0.369$\pm$0.002 &  -4.41 \\
HD20029 &  0.561 &   7.05 &        F7V &  0.156$\pm$0.001 &  -4.99 \\
HD20201 &  0.584 &   7.27 &        G0V &  0.178$\pm$0.001 &  -4.87 \\
HD20766 &  0.641 &   5.53 &        G2V &  0.270$\pm$0.002 &  -4.58 \\
*HD20782 &  0.630 &   7.36 &        G3V &  0.186$\pm$0.001 &  -4.85 \\
HD20807 &  0.600 &   5.24 &        G1V &  0.184$\pm$0.001 &  -4.84 \\
HD22104 &  0.679 &   8.32 &        G3V &  0.155$\pm$0.001 &  -5.04 \\
*HD23079 &  0.583 &   7.12 &     F8/G0V &  0.163$\pm$0.001 &  -4.95 \\
HD23127 &  0.690 &   8.58 &        G2V &  0.162$\pm$0.001 &  -5.00 \\
HD23484 &  0.870 &   6.99 &        K1V &  0.508$\pm$0.003 &  -4.43 \\
HD24112 &  0.560 &   7.24 &        F8V &  0.155$\pm$0.001 &  -5.00 \\
HD25587 &  0.543 &   7.40 &        F7V &  0.164$\pm$0.001 &  -4.93 \\
HD25874 &  0.667 &   6.74 &     G5IV-V &  0.169$\pm$0.001 &  -4.95 \\
HD26754 &  0.551 &   7.16 &     F7/F8V &  0.159$\pm$0.001 &  -4.97 \\
*HD27442 &  1.078 &   4.44 &       K2IV &  0.129$\pm$0.001 &  -5.35 \\
HD28255 &  0.659 &   6.28 &        G4V &  0.188$\pm$0.001 &  -4.85 \\
*HD30177 &  0.773 &   8.41 &        G8V &  0.152$\pm$0.001 &  -5.08 \\
HD30876 &  0.901 &   7.49 &        K2V &  0.468$\pm$0.003 &  -4.51 \\
HD31527 &  0.606 &   7.49 &        G2V &  0.172$\pm$0.001 &  -4.91 \\
HD36108 &  0.590 &   6.78 &        G3V &  0.155$\pm$0.001 &  -5.01 \\
HD38283 &  0.584 &   6.69 &     G0/G1V &  0.161$\pm$0.001 &  -4.97 \\
HD38382 &  0.580 &   6.34 &     F8/G0V &  0.167$\pm$0.001 &  -4.93 \\
HD38973 &  0.594 &   6.63 &        G2V &  0.161$\pm$0.001 &  -4.97 \\
HD40307 &  0.935 &   7.17 &        K3V &  0.265$\pm$0.002 &  -4.83 \\
HD202628 &  0.637 &   6.75 &        G5V &  0.257$\pm$0.002 &  -4.61 \\
HD204385 &  0.596 &   7.14 &       G0IV &  0.166$\pm$0.001 &  -4.94 \\
HD204961 &  1.521 &   8.66 &        M1V &  0.985$\pm$0.007 &  -5.10 \\
HD205390 &  0.879 &   7.14 &        K2V &  0.426$\pm$0.003 &  -4.53 \\
HD205536 &  0.755 &   7.07 &        G8V &  0.164$\pm$0.001 &  -5.01 \\
HD209268 &  0.564 &   6.88 &        F7V &  0.156$\pm$0.001 &  -4.99 \\
HD211317 &  0.650 &   7.26 &   G5III/IV &  0.156$\pm$0.001 &  -5.02 \\
HD212168 &  0.599 &   6.12 &       G3IV &  0.161$\pm$0.001 &  -4.97 \\
*HD216437 &  0.660 &   6.04 &     G4IV-V &  0.155$\pm$0.001 &  -5.03 \\
HD217987 &  1.483 &   7.35 &     M2/M3V &  1.081$\pm$0.007 &  -5.01 \\
HD222237 &  0.989 &   7.09 &        K3V &  0.296$\pm$0.002 &  -4.84 \\
HD222335 &  0.802 &   7.18 &        K1V &  0.238$\pm$0.002 &  -4.77 \\

 & & & & & \\
\underline{2002 July 20$^{\rmn{th}}$} &        &       &       &    &                         \\
GL551 &  1.807 &  11.01 &       M5Ve & 10.686$\pm$0.072 &  -4.28 \\
GL729 &  1.510 &  10.37 &     M3.5Ve &  6.188$\pm$0.042 &  -4.29 \\
HD1273 &  0.655 &   6.84 &        G2V &  0.192$\pm$0.001 &  -4.83 \\
HD1581 &  0.576 &   4.23 &        F9V &  0.167$\pm$0.001 &  -4.92 \\

\hline
\end{tabular} 
\medskip
\end{table*}

\begin{table*}
\begin{tabular}{cccccc}
\hline

HD2071 &  0.681 &   7.27 &    G8IV &  0.184$\pm$0.001 &  -4.88 \\
HD5133 &  0.936 &   7.15 &        K2V &  0.464$\pm$0.003 &  -4.56 \\
HD5562 &  0.808 &   7.17 &       G8IV &  0.147$\pm$0.001 &  -5.11 \\
HD7570 &  0.571 &   4.97 &        F8V &  0.159$\pm$0.001 &  -4.97 \\
HD7693 &  1.000 &   7.22 &        K2V &  0.655$\pm$0.004 &  -4.48 \\
HD8581 &  0.569 &   6.85 &        F8V &  0.152$\pm$0.001 &  -5.03 \\
HD9540 &  0.766 &   6.97 &        K0V &  0.365$\pm$0.002 &  -4.49 \\
HD12042 &  0.487 &   6.10 &        F8V &  0.161$\pm$0.001 &  -4.94 \\
HD13246 &  0.544 &   7.50 &        F8V &  0.341$\pm$0.002 &  -4.38 \\
HD17925 &  0.862 &   6.05 &        K1V &  0.674$\pm$0.004 &  -4.29 \\
*HD22049 &  0.881 &   3.72 &        K2V &  0.480$\pm$0.003 &  -4.47 \\
HD23456 &  0.511 &   6.97 &        G1V &  0.166$\pm$0.001 &  -4.91 \\
HD27274 &  1.115 &   7.64 &        K5V &  0.416$\pm$0.003 &  -4.86 \\
HD30295 &  0.812 &   8.86 &     K0/K1V &  0.142$\pm$0.001 &  -5.13 \\
HD31827 &  0.770 &   8.26 &       G8IV &  0.149$\pm$0.001 &  -5.09 \\
HD33811 &  0.765 &   8.71 &     G8IV/V &  0.164$\pm$0.001 &  -5.01 \\
HD38110 &  0.696 &   8.18 &         G5 &  0.159$\pm$0.001 &  -5.02 \\
HD38393 &  0.481 &   3.59 &        F7V &  0.167$\pm$0.001 &  -4.90 \\
HD40307 &  0.935 &   7.17 &        K3V &  0.255$\pm$0.002 &  -4.85 \\
HD84117 &  0.534 &   4.93 &        G0V &  0.158$\pm$0.001 &  -4.97 \\
HD85512 &  1.156 &   7.67 &        K5V &  0.312$\pm$0.002 &  -5.05 \\
HD85512 &  1.156 &   7.67 &        K5V &  0.383$\pm$0.002 &  -4.96 \\
HD85683 &  0.546 &   7.34 &        F8V &  0.165$\pm$0.001 &  -4.93 \\
HD85683 &  0.546 &   7.34 &        F8V &  0.168$\pm$0.001 &  -4.91 \\
HD85683 &  0.546 &   7.34 &        F8V &  0.168$\pm$0.001 &  -4.91 \\
HD97998 &  0.626 &   7.36 &        G5V &  0.187$\pm$0.001 &  -4.84 \\
HD101581 &  1.064 &   7.77 &        K5V &  0.500$\pm$0.003 &  -4.70 \\
HD101805 &  0.528 &   6.48 &        G1V &  0.160$\pm$0.001 &  -4.95 \\
*HD102117 &  0.721 &   7.47 &        G6V &  0.158$\pm$0.001 &  -5.03 \\
HD103026 &  0.554 &   5.85 &        F8V &  0.150$\pm$0.001 &  -5.04 \\
HD103493B &  0.646 &   6.70 &        G5V &  0.194$\pm$0.001 &  -4.82 \\
HD103975 &  0.527 &   6.76 &        G0V &  0.162$\pm$0.001 &  -4.94 \\
HD105328 &  0.613 &   6.72 &        G2V &  0.161$\pm$0.001 &  -4.97 \\
HD106453 &  0.711 &   7.47 &     K0/K1V &  0.319$\pm$0.002 &  -4.52 \\
HD106869 &  0.574 &   6.81 &        G1V &  0.162$\pm$0.001 &  -4.96 \\
HD110810 &  0.937 &   7.82 &        K3V &  0.623$\pm$0.004 &  -4.42 \\
HD112019 &  0.520 &   7.69 &        G0V &  0.165$\pm$0.001 &  -4.92 \\
HD113027 &  0.569 &   7.56 &        G2V &  0.189$\pm$0.001 &  -4.81 \\
HD114260 &  0.718 &   7.36 &        G6V &  0.173$\pm$0.001 &  -4.95 \\
HD114613 &  0.693 &   4.85 &        G3V &  0.157$\pm$0.001 &  -5.03 \\
HD114613 &  0.693 &   4.85 &        G3V &  0.157$\pm$0.001 &  -5.03 \\
HD115585 &  0.742 &   7.43 &     G6IV-V &  0.152$\pm$0.001 &  -5.07 \\
HD117105 &  0.583 &   7.20 &        G1V &  0.167$\pm$0.001 &  -4.93 \\
HD117939 &  0.669 &   7.29 &        G3V &  0.181$\pm$0.001 &  -4.89 \\
HD118475 &  0.618 &   6.97 &  G2/G3IV/V &  0.166$\pm$0.001 &  -4.95 \\
HD118972 &  0.855 &   6.92 &        K1V &  0.503$\pm$0.003 &  -4.42 \\
HD120780 &  0.891 &   7.37 &        K1V &  0.264$\pm$0.002 &  -4.79 \\
HD122862 &  0.581 &   6.02 &        G1V &  0.157$\pm$0.001 &  -4.99 \\
HD124584 &  0.590 &   7.29 &     G0/G1V &  0.160$\pm$0.001 &  -4.98 \\
HD125072 &  1.017 &   6.66 &        K3V &  0.283$\pm$0.002 &  -4.89 \\
HD125370 &  1.095 &   8.53 &      K0III &  0.179$\pm$0.001 &  -5.21 \\
HD128674 &  0.672 &   7.39 &        G5V &  0.188$\pm$0.001 &  -4.86 \\
HD129060 &  0.553 &   6.99 &        F7V &  0.330$\pm$0.002 &  -4.41 \\
HD134606 &  0.740 &   6.86 &       G5IV &  0.157$\pm$0.001 &  -5.04 \\
*HD134987 &  0.691 &   6.47 &        G5V &  0.155$\pm$0.001 &  -5.04 \\
HD136352 &  0.639 &   5.65 &        G2V &  0.176$\pm$0.001 &  -4.90 \\
HD140785 &  0.660 &   7.38 &        G5V &  0.154$\pm$0.001 &  -5.04 \\
HD140901 &  0.715 &   6.01 &       G6IV &  0.256$\pm$0.002 &  -4.66 \\
*HD142415 &  0.621 &   7.33 &        G1V &  0.235$\pm$0.002 &  -4.66 \\
HD143114 &  0.606 &   7.34 &        G3V &  0.171$\pm$0.001 &  -4.92 \\
HD144009 &  0.714 &   7.23 &        G8V &  0.193$\pm$0.001 &  -4.86 \\
HD144628 &  0.856 &   7.11 &        K3V &  0.193$\pm$0.001 &  -4.94 \\
HD145417 &  0.815 &   7.53 &        K0V &  0.210$\pm$0.001 &  -4.86 \\
HD145809 &  0.617 &   6.68 &        G3V &  0.154$\pm$0.001 &  -5.02 \\
HD146481 &  0.642 &   7.09 &        G4V &  0.160$\pm$0.001 &  -4.99 \\
HD147723 &  0.625 &   5.40 &       G0IV &  0.147$\pm$0.001 &  -5.08 \\
HD147723 &  0.625 &   5.40 &       G0IV &  0.159$\pm$0.001 &  -4.99 \\
HD149612 &  0.616 &   7.01 &        G3V &  0.179$\pm$0.001 &  -4.87 \\

\hline
\end{tabular} 
\medskip
\end{table*}

\begin{table*}
\begin{tabular}{cccccc}
\hline

HD150474 &  0.780 &   7.16 &        G8V &  0.148$\pm$0.001 &  -5.10 \\
HD151337 &  0.901 &   7.38 &        K0V &  0.147$\pm$0.001 &  -5.13 \\
HD152311 &  0.685 &   5.86 &       G5IV &  0.150$\pm$0.001 &  -5.07 \\
HD153075 &  0.581 &   6.99 &        G0V &  0.172$\pm$0.001 &  -4.90 \\
HD154577 &  0.889 &   7.38 &        K0V &  0.245$\pm$0.002 &  -4.82 \\
*HD154857 &  0.699 &   7.24 &        G5V &  0.154$\pm$0.001 &  -5.05 \\
HD155918 &  0.607 &   7.00 &        G2V &  0.175$\pm$0.001 &  -4.89 \\
HD155974 &  0.479 &   6.09 &        F6V &  0.159$\pm$0.001 &  -4.95 \\
HD156274 &  0.764 &   5.47 &        M0V &  0.176$\pm$0.001 &  -4.95 \\
HD157060 &  0.541 &   6.42 &        F8V &  0.161$\pm$0.001 &  -4.95 \\
HD159868 &  0.714 &   7.24 &        G5V &  0.171$\pm$0.001 &  -4.96 \\
HD162396 &  0.523 &   6.19 &        F8V &  0.155$\pm$0.001 &  -4.98 \\
HD162521 &  0.451 &   6.36 &        F8V &  0.227$\pm$0.001 &  -4.63 \\
HD163272 &  0.614 &   7.39 &     G2/G3V &  0.163$\pm$0.001 &  -4.96 \\
HD165269 &  0.611 &   7.29 &        G1V &  0.170$\pm$0.001 &  -4.92 \\
HD166553 &  0.599 &   7.27 &     G1/G2V &  0.157$\pm$0.001 &  -5.00 \\
HD168060 &  0.759 &   7.34 &        G5V &  0.148$\pm$0.001 &  -5.10 \\
*HD169830 &  0.517 &   5.90 &        F8V &  0.151$\pm$0.001 &  -5.02 \\
HD171990 &  0.593 &   6.39 &        G2V &  0.150$\pm$0.001 &  -5.05 \\
HD179140 &  0.627 &   7.23 &        G2V &  0.158$\pm$0.001 &  -5.00 \\
*HD179949 &  0.548 &   6.25 &        F8V &  0.198$\pm$0.001 &  -4.76 \\
HD183877 &  0.675 &   7.14 &      K0 &  0.183$\pm$0.001 &  -4.88 \\
HD184509 &  0.557 &   6.74 &        G1V &  0.174$\pm$0.001 &  -4.88 \\
HD188641 &  0.626 &   7.34 &        G2V &  0.159$\pm$0.001 &  -5.00 \\
HD190248 &  0.751 &   3.55 &     G5IV-V &  0.160$\pm$0.001 &  -5.03 \\
HD191408 &  0.868 &   5.32 &        K2V &  0.195$\pm$0.001 &  -4.94 \\
HD191849 &  1.431 &   7.97 &        M0V &  1.630$\pm$0.011 &  -4.76 \\
HD192310 &  0.878 &   5.73 &        K3V &  0.204$\pm$0.001 &  -4.92 \\
HD192865 &  0.558 &   6.91 &        F8V &  0.158$\pm$0.001 &  -4.97 \\
HD193193 &  0.594 &   7.20 &        G2V &  0.165$\pm$0.001 &  -4.95 \\
HD196068 &  0.640 &   7.18 &        G5V &  0.150$\pm$0.001 &  -5.06 \\
HD196378 &  0.544 &   5.11 &        F8V &  0.158$\pm$0.001 &  -4.97 \\
HD196390 &  0.626 &   7.33 &        G3V &  0.203$\pm$0.001 &  -4.78 \\
HD196800 &  0.607 &   7.21 &     G1/G2V &  0.162$\pm$0.001 &  -4.97 \\
HD202457 &  0.689 &   6.60 &        G5V &  0.161$\pm$0.001 &  -5.00 \\
HD202560 &  1.397 &   6.69 &     M1/M2V &  0.993$\pm$0.007 &  -4.93 \\
HD203985 &  0.876 &   7.49 &        K0V &  0.190$\pm$0.001 &  -4.96 \\
HD204287 &  0.663 &   7.33 &        G3V &  0.154$\pm$0.001 &  -5.04 \\
HD206395 &  0.559 &   6.67 &       G0IV &  0.162$\pm$0.001 &  -4.95 \\
HD209100 &  1.056 &   4.69 &        K5V &  0.450$\pm$0.003 &  -4.74 \\
HD210272 &  0.663 &   7.22 &        G3V &  0.152$\pm$0.001 &  -5.05 \\
HD213042 &  1.080 &   7.65 &        K4V &  0.325$\pm$0.002 &  -4.92 \\
*HD216435 &  0.621 &   6.03 &       G3IV &  0.157$\pm$0.001 &  -5.01 \\
HD216803 &  1.094 &   6.48 &       K4Vp &  0.856$\pm$0.006 &  -4.51 \\
HD219048 &  0.733 &   6.90 &        G5V &  0.161$\pm$0.001 &  -5.02 \\
HD221420 &  0.681 &   5.82 &        G2V &  0.142$\pm$0.001 &  -5.13 \\
HD222668 &  0.835 &   7.35 &     G8IV/V &  0.150$\pm$0.001 &  -5.09 \\
HD224619 &  0.741 &   7.47 &        G8V &  0.184$\pm$0.001 &  -4.91 \\
HD225213 &  1.462 &   8.56 &        M2V &  0.383$\pm$0.002 &  -5.43 \\

 & & & & & \\
\underline{2003 April 21$^{\rmn{st}}$} &        &       &       &   &  \\
HD4447 &  0.908 &   8.78 &         K0 &  0.157$\pm$0.001 &  -5.10 \\
HD30876 &  0.901 &   7.49 &        K2V &  0.393$\pm$0.003 &  -4.59 \\
HD42902 &  0.623 &   8.92 &     G2/G3V &  0.148$\pm$0.001 &  -5.07 \\
HD44821 &  0.663 &   7.37 &      K0/1V &  0.319$\pm$0.002 &  -4.49 \\
HD44821 &  0.663 &   7.37 &      K0/1V &  0.322$\pm$0.002 &  -4.48 \\
HD45701 &  0.660 &   6.45 &   G3III/IV &  0.157$\pm$0.001 &  -5.02 \\
HD52447 &  0.605 &   8.38 &        G0V &  0.148$\pm$0.001 &  -5.06 \\
HD52447 &  0.605 &   8.38 &        G0V &  0.149$\pm$0.001 &  -5.06 \\
HD55693 &  0.660 &   7.17 &        G1V &  0.164$\pm$0.001 &  -4.98 \\
HD55720 &  0.705 &   7.50 &        G6V &  0.166$\pm$0.001 &  -4.98 \\
HD56560 &  0.737 &   7.33 &     G6IV/V &  0.143$\pm$0.001 &  -5.12 \\
HD59468 &  0.694 &   6.72 &     G5IV-V &  0.162$\pm$0.001 &  -5.00 \\
HD61686 &  0.693 &   8.54 &        G3V &  0.136$\pm$0.001 &  -5.18 \\
HD61686 &  0.693 &   8.54 &        G3V &  0.144$\pm$0.001 &  -5.12 \\
HD65907A &  0.573 &   5.59 &        G2V &  0.173$\pm$0.001 &  -4.89 \\
HD67199 &  0.872 &   7.18 &        K1V &  0.336$\pm$0.002 &  -4.64 \\
HD67556 &  0.548 &   7.30 &        F8V &  0.175$\pm$0.001 &  -4.87 \\

\hline
\end{tabular} 
\medskip
\end{table*}

\begin{table*}
\begin{tabular}{cccccc}
\hline

HD69655 &  0.579 &   6.63 &        G1V &  0.165$\pm$0.001 &  -4.94 \\
*HD70642 &  0.692 &   7.17 &     G8:III &  0.167$\pm$0.001 &  -4.97 \\
*HD70642 &  0.692 &   7.17 &     G8:III &  0.168$\pm$0.001 &  -4.97 \\
HD70889 &  0.600 &   7.09 &        G0V &  0.237$\pm$0.002 &  -4.64 \\
HD72769 &  0.745 &   7.22 &       K1IV &  0.147$\pm$0.001 &  -5.10 \\
HD73121 &  0.578 &   6.44 &        G1V &  0.169$\pm$0.001 &  -4.92 \\
HD74868 &  0.567 &   6.56 &       G3IV &  0.159$\pm$0.001 &  -4.98 \\
*HD76700 &  0.745 &   8.16 &        G8V &  0.130$\pm$0.001 &  -5.22 \\
*HD76700 &  0.745 &   8.16 &        G8V &  0.136$\pm$0.001 &  -5.18 \\
HD78429 &  0.664 &   7.31 &        G5V &  0.202$\pm$0.001 &  -4.80 \\
HD80635 &  0.729 &   8.80 &       G3IV &  0.135$\pm$0.001 &  -5.18 \\
HD80635 &  0.729 &   8.80 &       G3IV &  0.141$\pm$0.001 &  -5.14 \\
HD80913 &  0.556 &   7.49 &        F6V &  0.152$\pm$0.001 &  -5.02 \\
HD88201 &  0.558 &   7.45 &        G0V &  0.357$\pm$0.002 &  -4.36 \\
HD90712 &  0.585 &   7.52 &     G2/G3V &  0.388$\pm$0.003 &  -4.32 \\
HD109200 &  0.836 &   7.13 &        K0V &  0.189$\pm$0.001 &  -4.94 \\

 & & & & & \\
\underline{2004 August 23$^{\rmn{rd}}$ \& 24$^{\rmn{th}}$} &        &       &       &                         \\
HD7442	&	0.587	&	7.17	&	F8/G0V	&	0.151$\pm$0.001	&	-5.04	\\
HD13578	&	0.620	&	7.50	&	G3IV	&	0.156$\pm$0.001	&	-5.01	\\
HD16427	&	0.568	&	6.84	&	F8V	&	0.164$\pm$0.001	&	-4.94	\\
HD17925	&	0.862	&	6.05	&	K1V	&	0.699$\pm$0.005	&	-4.27	\\
HD21626	&	0.501	&	6.73	&	G0IV	&	0.147$\pm$0.001	&	-5.05	\\
*HD22049	&	0.881	&	3.72	&	K2V	&	0.563$\pm$0.004	&	-4.39	\\
HD22924	&	0.552	&	6.94	&	F8/G0V	&	0.152$\pm$0.001	&	-5.02	\\
HD23308	&	0.522	&	6.50	&	F8V	&	0.250$\pm$0.002	&	-4.57	\\
HD28454	&	0.470	&	6.10	&	F8V	&	0.160$\pm$0.001	&	-4.95	\\
HD31975	&	0.521	&	6.28	&	F8V	&	0.169$\pm$0.001	&	-4.90	\\
HD32820	&	0.528	&	6.30	&	F8V	&	0.159$\pm$0.001	&	-4.96	\\
HD33473	&	0.662	&	6.75	&	G3V	&	0.162$\pm$0.001	&	-4.99	\\
HD34606	&	1.014	&	8.92	&	G5	&	0.168$\pm$0.001	&	-5.14	\\
HD38393	&	0.481	&	3.59	&	F7V	&	0.158$\pm$0.001	&	-4.96	\\
*HD142415	&	0.621	&	7.33	&	G1V	&	0.260$\pm$0.002	&	-4.59	\\
HD146481	&	0.642	&	7.09	&	G4V	&	0.181$\pm$0.001	&	-4.88	\\
HD147722	&	0.625	&	5.40	&	G0	&	0.168$\pm$0.001	&	-4.94	\\
HD147722	&	0.625	&	5.40	&	G0	&	0.181$\pm$0.001	&	-4.87	\\
HD149612	&	0.616	&	7.01	&	G3V	&	0.205$\pm$0.001	&	-4.76	\\
HD150474	&	0.780	&	7.16	&	G8V	&	0.166$\pm$0.001	&	-5.00	\\
HD151337	&	0.901	&	7.38	&	K0V	&	0.158$\pm$0.001	&	-5.09	\\
HD153075	&	0.581	&	6.99	&	G0V	&	0.195$\pm$0.001	&	-4.79	\\
HD154577	&	0.889	&	7.38	&	K0V	&	0.252$\pm$0.002	&	-4.81	\\
*HD154857	&	0.699	&	7.24	&	G5V	&	0.173$\pm$0.001	&	-4.94	\\
HD155974	&	0.479	&	6.09	&	F6V	&	0.162$\pm$0.001	&	-4.93	\\
HD179140	&	0.627	&	7.23	&	G2V	&	0.175$\pm$0.001	&	-4.90	\\
*HD179949	&	0.548	&	6.25	&	F8V	&	0.209$\pm$0.001	&	-4.72	\\
HD183877	&	0.675	&	7.14	&	K0	&	0.194$\pm$0.001	&	-4.84	\\
HD188641	&	0.626	&	7.34	&	G2V	&	0.171$\pm$0.001	&	-4.92	\\
HD192310	&	0.878	&	5.73	&	K3V	&	0.238$\pm$0.002	&	-4.83	\\
HD192865	&	0.558	&	6.91	&	F8V	&	0.164$\pm$0.001	&	-4.94	\\
HD193193	&	0.594	&	7.20	&	G2V	&	0.186$\pm$0.001	&	-4.83	\\
HD196068	&	0.640	&	7.18	&	G5V	&	0.160$\pm$0.001	&	-5.00	\\
HD196378	&	0.544	&	5.11	&	F8V	&	0.162$\pm$0.001	&	-4.95	\\
HD196390	&	0.626	&	7.33	&	G3V	&	0.243$\pm$0.002	&	-4.64	\\
HD196800	&	0.607	&	7.21	&	G1/G2V	&	0.160$\pm$0.001	&	-4.98	\\
HD202560	&	1.397	&	6.69	&	M1/M2V	&	1.076$\pm$0.007	&	-4.89	\\
HD202560	&	1.397	&	6.69	&	M1/M2V	&	1.111$\pm$0.007	&	-4.88	\\
HD203985	&	0.876	&	7.49	&	K0V	&	0.188$\pm$0.001	&	-4.96	\\
HD203985	&	0.876	&	7.49	&	K0V	&	0.191$\pm$0.001	&	-4.95	\\
HD204287	&	0.663	&	7.33	&	G3V	&	0.164$\pm$0.001	&	-4.97	\\
HD207129	&	0.601	&	5.57	&	G2V	&	0.184$\pm$0.001	&	-4.85	\\
HD212330	&	0.665	&	5.31	&	F9V	&	0.164$\pm$0.001	&	-4.98	\\
*HD216435	&	0.621	&	6.03	&	G3IV	&	0.164$\pm$0.001	&	-4.96	\\
HD216803	&	1.094	&	6.48	&	K4Vp	&	1.084$\pm$0.007	&	-4.41	\\
HD219048	&	0.733	&	6.90	&	G5V	&	0.162$\pm$0.001	&	-5.01	\\
HD222668	&	0.835	&	7.35	&	G8IV/V	&	0.160$\pm$0.001	&	-5.05	\\
HD224619	&	0.741	&	7.47	&	G8V	&	0.192$\pm$0.001	&	-4.88	\\

\hline
\end{tabular} 
\medskip
\end{table*}

\begin{table*}
\begin{tabular}{cccccc}
\hline

 & & & & & \\
\underline{2005 June 16$^{\rmn{th}}$} &        &       &    &   &                        \\

HD56957 &  0.701 &   7.57 &        G3V &  0.148$\pm$0.001 &  -5.09 \\
HD63685 &  0.758 &   7.38 &        G5V &  0.157$\pm$0.001 &  -5.04 \\
HD67556 &  0.548 &   7.30 &        F8V &  0.196$\pm$0.001 &  -4.77 \\
HD69655 &  0.579 &   6.63 &        G1V &  0.176$\pm$0.001 &  -4.88 \\
*HD70642 &  0.692 &   7.17 &     G8:III &  0.175$\pm$0.001 &  -4.93 \\
HD72769 &  0.745 &   7.22 &       K1IV &  0.161$\pm$0.001 &  -5.02 \\
*HD73526 &  0.737 &   8.99 &        G6V &  0.154$\pm$0.001 &  -5.05 \\
HD74868 &  0.567 &   6.56 &       G3IV &  0.165$\pm$0.001 &  -4.93 \\
*HD76700 &  0.745 &   8.16 &        G8V &  0.160$\pm$0.001 &  -5.02 \\
HD80913 &  0.556 &   7.49 &        F6V &  0.166$\pm$0.001 &  -4.93 \\
HD94340 &  0.645 &   7.02 &     G3/G5V &  0.293$\pm$0.002 &  -4.53 \\
HD95456 &  0.527 &   6.06 &        F8V &  0.168$\pm$0.001 &  -4.91 \\
HD142022 &  0.790 &   7.70 &        K0V &  0.166$\pm$0.001 &  -5.01 \\

\hline
\end{tabular} 
\medskip
\end{table*}

\chapter{\bf{Acknowledgements}}

We acknowledge the comments made by the anonymous referee.

\bibliographystyle{mn2e}
\bibliography{refs}

\label{lastpage}

\end{document}